\documentclass[prl,reprint,twocolumn,superscriptaddress,floatfix,showpacs,10pt]{revtex4-1}%
\usepackage{amsfonts,amsmath,amssymb}
\usepackage[colorlinks=true,urlcolor=blue,anchorcolor=blue,citecolor=blue,filecolor=blue,linkcolor=blue,menucolor=blue,pagecolor=blue,linktocpage=true,pdfproducer=medialab,pdfa=true]{hyperref}
\usepackage{tikz}

\newcommand{\pb}[1] {\overline{#1}}
\newcommand{\p}[1] {\underline{#1}}
\newcommand{\dd}{\mathrm{d}}

\begin{document}

\title{\texorpdfstring{$\alpha'$}{alpha'}-corrected Poisson-Lie T-duality}

\author{Falk Hassler}
\email{falk@fhassler.de}
\affiliation{George P. \& Cynthia Woods Mitchell Institute for Fundamental Physics and Astronomy,\\
Texas A\&M University, College Station, TX 77843, USA}

\author{Thomas Rochais}
\email{thb@sas.upenn.edu}
\affiliation{Department of Physics and Astronomy, University of Pennsylvania, Philadelphia, PA 19104, USA}

\begin{abstract}
We propose leading order $\alpha'$-corrections to the Poisson-Lie T-duality transformation rules of the metric, $B$-field, and dilaton. Based on Double Field Theory, whose corrections to this order are known, we argue that they map conformal field theories to conformal field theories. Remarkably, Born geometry plays a central role in the construction.
\end{abstract}

\maketitle

Abelian T-duality is an important cornerstone in the framework of string theory. It is applicable to target space geometries that possess abelian isometries and a natural question is if it is possible to extend T-duality to more general situations. Non-abelian T-duality \cite{delaOssa:1992vci} arose from this idea and is based on the observation that the Buscher procedure \cite{Buscher:1987sk}, which describes abelian T-duality in the closed string $\sigma$-model, can be applied to non-abelian isometries, too. However, there are two major obstacles compared to the abelian case. First, the dual background has a smaller isometry group than the original one. Hence, it seems in general impossible to invert the transformation which is crucial to have a duality. Second, it is problematic to extract global properties of the dual target space. They are for example required to construct an operator mapping on higher genus Riemann surfaces \cite{Alvarez:1993qi}. Poisson-Lie (PL) T-duality arises from an elegant solution to the first problem. It is based on the seminal observation \cite{Klimcik:1995ux} that both $\sigma$-models, which describe either the target space or its dual, originate from the same structure, a Drinfeld double. It governs the Hamiltonian dynamics of the models, and their equivalence is guaranteed by a canonical transformation. Remarkably, non-abelian T-duality constraints the Drinfeld double significantly. But the idea implemented in \cite{Klimcik:1995ux} works as well without this restriction. Hence, PL T-duality provides a more general notion of T-duality whose name originates from the fact that it relates target spaces that are PL groups. Like abelian and non-abelian T-duality are only applicable to target space geometries with isometries, a related notion exists for PL T-duality. It is based on non-commutative conserved currents on the worldsheet \cite{Klimcik:1995ux} which generate PL symmetry. Despite their intriguing mathematical structure and physical properties, research activity in $\sigma$-models with PL symmetric target spaces was moderate for almost two decades. Most arguably because they inherit the problems on global properties that non-abelian T-duality already faces. Just six years ago, when their relation to integrable string worldsheet theories was fully appreciated \cite{Delduc:2013qra}, significant new interest arose. Due to the astonishing success with which integrability was applied in the AdS/CFT correspondence to explore 4D maximally supersymmetric Yang-Mills theory in the large $N$ limit beyond the perturbative regime \cite{Beisert:2010jr}, the demand for new integrable $\sigma$-models is high and a vast new field of applications opens up for PL symmetry and T-duality. 

In this context, a particularly important question is how PL T-duality is affected by quantum corrections. They are controlled in string theory by two parameters: $\alpha'$ and $g_\mathrm{S}$. The former captures the extended nature of the string and the latter its ability to split. Abelian T-duality is a genuine symmetry of string theory and therefore applies to all orders in $\alpha'$ and $g_\mathrm{S}$ \cite{Rocek:1991ps}. For PL T-duality the situation is more subtle. Because of the notorious problem with higher genus worldsheets, there is currently not much to say about the fate of $g_\mathrm{S}$-corrections. However, this does not rule out the possibility of extending the validity of PL T-duality beyond the leading order in $\alpha'$. On the contrary, recently computed $\alpha'$-corrections of integrable deformations point very clearly in this direction \cite{Hoare:2019ark,*Hoare:2019mcc,Borsato:2020bqo}. Hence, the objective of this letter is to construct leading order $\alpha'$-corrections to the PL T-duality transformation rules in a bosonic $\sigma$-model and to argue that they preserve conformal invariance. Key to this endeavour are three techniques: The formulation of PL symmetric target space geometries in the framework of Double Field Theory (DFT) \cite{Hassler:2017yza}, the $\alpha'$-corrected DFT flux formulation introduced by Marqu\'es and Nu$\tilde{\mathrm{n}}$ez \cite{Marques:2015vua}, and finite generalised Green-Schwarz (gGS) transformations recently presented by Borsato, L{\'o}pez, and Wulff \cite{Borsato:2020bqo}.

\paragraph*{PL T-duality and DFT:} Directly at the level of the metric, $B$-field, and dilaton, PL symmetric target spaces might look very complicated. But fortunately, their underlying structure becomes much simpler in the framework of DFT \cite{Siegel:1993th,*Hull:2009mi,*Hohm:2010pp}, where they are expressed in the language of generalised geometry. More precisely, the metric and the $B$-field can be unified in a generalised frame field \cite{Hohm:2010xe} $E_A{}^I$ on the generalised tangent space. It is governed by the frame algebra \cite{Hassler:2017yza}
\begin{equation}\label{eqn:framealg}
  \mathcal{L}_{E_A} E_B{}^I = F_{AB}{}^C E_C{}^I
\end{equation}
where $\mathcal{L}$ denotes the generalised Lie derivative
\begin{equation}
  \mathcal{L}_{E_A} E_B{}^I = E_A{}^J \partial_J E_B{}^I + \big( \partial^I E_{AJ} - \partial_J E_A{}^I \big) E_B{}^J
\end{equation}
and $F_{AB}{}^C$ are the structure constants of a Lie algebra $\mathfrak{g}$, generating the corresponding Lie group $G$. Uppercase, Latin characters denote doubled indices, running from $1,\dots,2 D$. They come in two different kinds: flat indices ranging from $A$ to $H$ and curved indices starting with $I$. Both are related by the generalised frame field. They are raised and lowered with
\begin{equation}
  \eta^{IJ} = \begin{pmatrix}
    0 & \delta_i{}^j \\
    \delta^i{}_j & 0
  \end{pmatrix}\,, \quad
  \eta^{AB} = \begin{pmatrix}
    \eta^{ab} & 0 \\
    0 & -\eta^{\bar a\bar b}
  \end{pmatrix}\,,
\end{equation}
and their respective inverses, where $\eta^{ab}=\eta^{\bar a\bar b}$ has either Lorentzian or Euclidean signature. Furthermore, we always deal with the canonical solution to the section condition $\partial_I = ( 0 \,\, \partial_i )$.

Frame fields $E_A{}^I$ that satisfy~\eqref{eqn:framealg}, can be constructed systematically on the coset $H \backslash G$, if $H$ is a maximally isotropic subgroup of $G$ \cite{Hassler:2019wvn,*Demulder:2018lmj}. Isotropy is defined in terms of an O($D$,$D$) invariant pairing $\langle \cdot \,,\, \cdot \rangle$ on $\mathfrak{g}$. It is equivalent to $\eta_{AB}$, once an appropriate set of $2D$ linearly independent generators $t_A \in \mathfrak{g}$ is chosen. In this case, we identify $\langle t_A, t_B \rangle = \eta_{AB}$ and define a maximally isotropic subgroup $H$ as a subgroup of $G$ which has the maximal number of linearly independent generators that are pairwise annihilated by the pairing. Taking into account the signature of $\eta_{AB}$, it follows that $\dim H = D$. Depending on $G$ and the pairing, different subgroups (labeled $H_1$, $H_2$, \dots) might have this property. This observation is directly related to PL T-duality because each of them results in a generalised frame field describing a different, but still physically equivalent, target space geometry. At this point, the term duality is slightly misleading because it implies that there are at most $H_1$ and $H_2$ which is not true. Thus in general, one might prefer to refer to PL plurality. There are two major ingredients that enter the construction of the generalised frame field. The right-action of $G$ on the coset $H\backslash G$ gives rise to $2 D$ vector fields $k_A{}^i \partial_i$. They furnish the frame algebra
\begin{equation}
  L_{k_A} k_B{}^i = F_{AB}{}^C k_C{}^i
\end{equation}
under the standard Lie derivative $L$. It matches the vector part of \eqref{eqn:framealg} and therefore it is natural to identify $E_A{}^i = k_A{}^i$. To complete the construction, we also need the corresponding one form part \cite{Hassler:2019wvn}
\begin{equation}
  E_{Ai} \dd x^i = \langle t_A, l \rangle - \frac12\langle \iota_{k_A} l, l\rangle - \iota_{k_A} B_\mathrm{WZW}
\end{equation}
with $l=m^{-1} \dd m$, $m \in H \backslash G$. In general, it contains a locally defined $B$-field which captures the WZW-term of the underlying $\sigma$-model
\begin{equation}
  \dd B_\mathrm{WZW} = \frac1{3!} \langle l \overset{\wedge}{,} l \wedge l \rangle \,.
\end{equation}

For latter convenience, we parameterise the result in terms of three quantities: the frame field $e_a{}^i$ whose inverse transpose is denoted by $e^a{}_i$, the $B$-field $B_{ij}$ and a Lorentz transformation $\Lambda_{\bar a}{}^{\bar b}$ with the defining property $\Lambda_{\bar a}{}^{\bar c} \Lambda_{\bar b}{}^{\bar d} \eta_{\bar c\bar d} = \eta_{\bar a\bar b}$,
\begin{equation}\label{eqn:decompframe}
  E_A{}^I = \frac{1}{\sqrt{2}} \begin{pmatrix}
    \delta_a{}^b & 0 \\
    0 & \Lambda_{\bar a}{}^{\bar b}
  \end{pmatrix}\begin{pmatrix}
    e_{bi} + e_b{}^j B_{ji} &  e_b{}^i \\
    - e_{\bar bi} + e_{\bar b}{}^j B_{ji} & e_{\bar b}{}^i
  \end{pmatrix}\,.
\end{equation}
While $e^a{}_i$=$e^{\bar a}{}_i$, which gives rise to the metric $g_{ij}$=$e^a{}_i e^b{}_j \eta_{ab}$, and $B_{ij}$ shape the target space directly, the role of $\Lambda_{\bar a}{}^{\bar b}$ is more subtle. In a bosonic $\sigma$-model at the classical level, it is irrelevant. Still, it is crucial for \eqref{eqn:framealg} to hold and we will see that it plays a significant role for $\alpha'$-corrections to PL T-duality. Remarkably, the same is true for the R/R sector of type II superstrings where $\Lambda_{\bar a}{}^{\bar b}$ already affects the transformation rules to leading order $\alpha'$ \cite{Hassler:2017yza}. Similar to abelian T-duality, the PL T-duality transformation rules for the metric and $B$-field can be elegantly written in terms of the generalised metric
\begin{equation}\label{eqn:genmetric}
  \mathcal{H}^{IJ} = \begin{pmatrix} g_{ij} - B_{ik} g^{kl} B_{lj} & -B_{ik} g^{kj} \\
    g^{ik} B_{kj} & g^{ij}
    \end{pmatrix}
\end{equation}
as a coordinate dependent O($D$,$D$) transformation \cite{Hassler:2017yza}. We construct the latter by assuming that $G$ has at least two different maximally isotropic subgroups $H$ and $\widetilde H$ because only then PL T-duality is applicable. For both, we construct the generalised frame fields, $E_A{}^I$ and $\widetilde{E}_A{}^I$, to eventually extract
\begin{equation}\label{eqn:defO}
  O_I{}^J = E^A{}_I \widetilde{E}_A{}^J\,.
\end{equation}
It mediates the O($D$,$D$) transformation which relates both PL T-dual backgrounds,
\begin{equation}\label{eqn:trgenmetric0}
  \widetilde{\mathcal{H}}^{IJ} = O_K{}^I O_L{}^J \mathcal{H}^{KL}\,.
\end{equation}

A huge advantage of this approach is that it emphasises the invariance of $F_{AB}{}^C$ in \eqref{eqn:framealg} under PL T-duality. Furthermore, the flux formulation of DFT \cite{Geissbuhler:2013uka} allows us to rewrite the low-energy effective action
\begin{equation}\label{eqn:Seff}
  S = \int\dd^{D}x\,
    \sqrt{g} e^{-2\phi} \big(R + 4 (\partial \phi)^2 
    - \frac{1}{12} H^2 \big)
\end{equation}
and its field equations exclusively in terms of $F_{ABC}$ and
\begin{equation}
  F_A = 2 E_A{}^I \partial_I d + E^{BI} \partial_I E_B{}^J E_{AJ}
\end{equation}
with the generalised dilaton $d = \phi - \frac12 \log \sqrt{g}$. It is natural to assume that since $F_{ABC}$ is invariant under PL T-duality, $F_A$ should be, too. Imposing this additional constraint fixes the transformation of the generalised dilaton
\begin{equation}\label{eqn:trgend}
  \partial_I \widetilde{d} = O^J{}_I \partial_J d + \frac12 O_I
\end{equation}
with
\begin{equation}
  O_I = \partial_J \left( \widetilde{E}_A{}^J - E_A{}^J \right) E^A{}_I\,.
\end{equation}
Because $O_I{}^J$ and $O_I$ depend simultaneously on the coordinates of $H \backslash G$ and of its dual $\widetilde{H} \backslash G$, one might be worried to end up with target space fields that depend on unphysical coordinates after the transformation. Fortunately, for PL symmetric target spaces this situation is ruled out. It is common lore that it can be very hard to spot this symmetry directly at the level of the target space fields. Hence, it is more common to start from the doubled description in terms of the Lie group $G$ combined with the constants $F_{ABC}$, $F_A$ and then extract both PL T-dual target spaces according to the diagram
\begin{center}
  \begin{tikzpicture}
    \node[draw,rectangle,name=G] {$G$, $F_{ABC}$, $F_A$};
    \node[at=(G.south west),anchor=north east,rectangle,draw,name=G/H,xshift=-1em,yshift=-1em] {%
      $H\backslash G$, $\mathcal{H}^{IJ}$, $d$};
    \node[at=(G.south east),anchor=north west,rectangle,draw,name=G/tH,xshift=+1em,yshift=-1em] {%
      $\widetilde{H}\backslash G$, $\widetilde{\mathcal{H}}^{IJ}$, $\widetilde{d}$};
    \draw[<->] (G/H.east) -- (G/tH.west) node[midway,below] {$O_I{}^J$, $O_I$};
    \draw[<->] (G.west) -- (G/H.north)  node[midway,xshift=-1em,yshift=1em] {$E_A{}^I$};
    \draw[<->] (G.east) -- (G/tH.north) node[midway,xshift=+1em,yshift=1em] {$\widetilde{E}_A{}^I$};
    \node[at=(G/tH.south east),anchor=south west] {.};
  \end{tikzpicture}
\end{center}

Particularly interesting are target space geometries whose metric, $B$-field and dilaton solve the field equations of the effective action \eqref{eqn:Seff} because they give rise to conformal field theories (CFTs) on the worldsheet (at least at the one-loop level). Hence, we conclude that because the field equations do not change under PL T-duality, solutions are mapped to solutions and therefore conformal invariance is preserved. At one loop this statement can be further refined. A CFT can be perturbed by a relevant deformation which triggers an RG flow from the UV to either another CFT or a gapped phase in the IR. PL symmetric $\sigma$-models are one-loop renormalisable \cite{Valent:2009nv} and again, their $\beta$-functions can be expressed exclusively in terms of  $F_{ABC}$ and $F_A$ \cite{Sfetsos:2009vt}. Hence, PL T-duality does not only preserve fixed points but rather the complete RG flow.

\paragraph*{$\alpha'$-corrected DFT:} We will now show how this argumentation extends beyond the leading order of $\alpha'$. A major challenge is that beyond one loop, all relevant quantities like the effective action or $\beta$-functions become renormalisation scheme dependent. Different schemes are related by field redefinitions. Eventually, this dependence drops out for physical observables but during all intermediate steps, it is essential to keep track of it. Consequentially, there is no universal expression for the four-derivative effective action comparable to ~\eqref{eqn:framealg}, but rather one for every scheme. Popular schemes are the Metsaev-Tseytlin (MT) \cite{Metsaev:1987zx}, Hull-Townsend (HT) \cite{Hull:1987yi} and the generalised Bergshoeff-de Roo (gBR) \cite{Bergshoeff:1989de,Marques:2015vua} scheme. Choosing an appropriate scheme can simplify calculations significantly. In particular, it affects how symmetries of the theory are realised. An example is that while in the MT or HT scheme the action of diffeomorphisms and Lorentz transformations is the same as at one loop, the $B$-field Lorentz transformations in the gBR scheme receive a correction. Intriguingly, this correction is required to facilitate the Green-Schwarz (GS) anomaly cancellation mechanism for the heterotic superstring.

Because~\eqref{eqn:framealg} has proven to be a fundamental identity for all PL symmetric backgrounds, we prefer a scheme where it still holds unchanged. Furthermore, the effective action in this scheme should be exclusively captured by $F_{AB}{}^C$ and $F_A$ like before. Fortunately, a scheme with exactly these properties exists \cite{Marques:2015vua,Baron:2017dvb} and we will refer to it as Marqu\'es-Nu\~{n}ez (MN) scheme. While not affecting generalised diffeomorphisms, which is essential to keeping the construction of generalised frame fields from above applicable, it modifies double Lorentz transformations. At leading order, the latter leave by definition the generalised metric invariant. This is the reason why we could safely ignore $\Lambda_{\bar a}{}^{\bar b}$ in \eqref{eqn:trgenmetric0}. Beyond that order, it has to be included and results in $\alpha'$ corrected transformation rules. More precisely, except for $\eta_{IJ}$, all quantities will receive $\alpha'$-corrections. They are labelled by $\mathcal{H}_{IJ} = \mathcal{H}^{(0)}_{IJ} + \alpha' \mathcal{H}^{(1)}_{IJ} + \mathcal{O}(\alpha'^2)$. Finite double Lorentz transformations, also called gGS transformations, are denoted by $\mathcal{H}_{IJ} \rightarrow \mathcal{H}_{IJ} + \Delta_\Lambda \mathcal{H}_{IJ}$ where 
\begin{equation}
  \Lambda_A{}^B = \begin{pmatrix}
    \Lambda_a{}^b & 0 \\
    0 & \Lambda_{\bar a}{}^{\bar b}
  \end{pmatrix}
\end{equation}
is the parameter of the transformation. For our purpose, it is sufficient to restrict it to the form of the first matrix in the generalised frame field~\eqref{eqn:decompframe} and thus set $\Lambda_a{}^b = \delta_a{}^b$. A major challenge is that \cite{Marques:2015vua} does not present finite gGS transformations $\Delta_\Lambda$, but only the infinitesimal version $\delta_\lambda$, with $\Lambda = \exp(\lambda)$. While it should be possible to formally integrate $\delta_\lambda$, we find it more convenient to make an educated guess of how a finite counterpart might look like and then show that it is compatible with the infinitesimal transformations in \cite{Marques:2015vua}. A similar approach allowed \cite{Borsato:2020bqo} to present finite gGS transformations for the metric and $B$-field. However, at this level, the elegant structure of PL T-duality is not manifest. Hence, we prefer to discuss doubled quantities, like the generalised metric or dilaton. Remarkably, their transformation cannot be written exclusively in terms of $\eta_{IJ}$, $\mathcal{H}_{IJ}$, $E_A{}^J$ and $F_{ABC}$. It additionally depends on an involution $K_I{}^J$, with the leading contribution
\begin{equation}
  K^{(0)}_I{}^J = \begin{pmatrix}
    - \delta^i{}_j & 0 \\
    2 B^{(0)}_{ij} & \delta_i{}^j
  \end{pmatrix}\,,
\end{equation}
which equips the target space with an almost Born structure \cite{Freidel:2013zga} (we do not require its integrability).

\paragraph*{Finite gGS transformations and PL T-duality:} It is this structure which eventually facilitates to write down a proposal for the finite gGS transformation of the generalised metric
\begin{equation}\label{eqn:doubleLgenmetric}
  \Delta_\Lambda^{(1)} \mathcal{H}_{IJ} = K^{(0)}_{(I|}{}^K \left( \mathcal{H}^{(0)}_{KL} \Delta^{(1)}_\Lambda K_{|J)}{}^L + 2 \Delta^{(0)}_\Lambda \mathcal{G}_{|J)K} \right)
\end{equation}
with
\begin{equation}
  \mathcal{G}^{(0)}_{IJ} = - \frac12 F^{(0)}_{\pb I\p A}{}^{\p B} F^{(0)}_{\pb J\p B}{}^{\p A}\,.
\end{equation}
Here, we adopt the notation of \cite{Marques:2015vua} to indicate indices that are projected by either $P_{IJ}= \frac12 ( \eta_{IJ} - \mathcal{H}_{IJ} )$ as $V_{\p I} = P_I{}^J V_J$ or $\overline{P}_{IJ} = \frac12 ( \eta_{IJ} + \mathcal{H}_{IJ} )$ as $V_{\pb I} = \overline{P}_I{}^J V_J$. Taking into account that the generalised frame field transforms to leading order as $E_A{}^I \rightarrow \Lambda_A{}^B E_B{}^I$, it is straightforward to obtain
\begin{equation}\label{eqn:trafoG}
  \Delta^{(0)}_\Lambda \mathcal{G}_{IJ} = F^{(0)}_{(\pb I|\p A}{}^{\p B} \Theta_{|\pb J)\p B}{}^{\p A} - \frac12 \Theta_{\pb I\p A}{}^{\p B} \Theta_{\pb J\p B}{}^{\p A}
\end{equation}
where $\Theta_{IA}{}^B$ captures the left-invariant Maurer-Cartan form (the invariant left-action is $\Lambda_A{}^B \rightarrow \Lambda'_A{}^C \Lambda_C{}^B$ where $\Lambda'_A{}^B$ is constant)
\begin{equation}
  \Theta_{IA}{}^B = \partial_I \Lambda^C{}_A \Lambda_C{}^B
\end{equation}
with the corresponding Maurer-Cartan equation $2 \partial_{[I} \Theta_{J]A}{}^B = [ \Theta_I, \Theta_J ]_A{}^B$. Note that the proposed transformation \eqref{eqn:doubleLgenmetric} guarantees that the algebraic relations of the Born structure are preserved at order $\alpha'$.  To explicitly evaluate it, we additionally impose
\begin{equation}\label{eqn:doubleLK}
  \Delta^{(1)}_\Lambda K_{IJ} = \Delta^{(0)}_\Lambda \mathcal{B}_{IJ} - K^{(0)}_I{}^K \Delta^{(0)} \mathcal{B}_{KL} K^{(0)}_J{}^L
\end{equation}
with
\begin{equation}\label{eqn:trafoB}
  \Delta^{(0)}_\Lambda \mathcal{B}_{IJ} = F^{(0)}_{[\pb I|\p A}{}^{\p B} \Theta_{|\pb J]\p B}{}^{\p A} + \mathcal{B}_{\pb I\pb J}^{\mathrm{WZW}}
\end{equation}
and
\begin{equation}
  3 \partial_{[I}\mathcal{B}_{JK]}^{\mathrm{WZW}} = \Theta_{[I|A}{}^B \Theta_{|J|B}{}^C \Theta_{|K]C}{}^A \,.
\end{equation}

Eventually, we have to show that our proposal for finite gGS transformations is compatible with the known infinitesimal results mediated by $\delta_\lambda$ with the antisymmetric parameter $\lambda_{AB}$. In order to extract the latter from the former, the finite transformations are perturbed by the right action $\Lambda\rightarrow  \Lambda + \Lambda\lambda$, which only affects
\begin{equation}
  \tilde{\delta}_\lambda \Theta_{IA}{}^B = - \partial_I \lambda_{A}{}^{B} - \lambda_A{}^C \Theta_{IC}{}^B - \Theta_{IA}{}^C \lambda^B{}_C\,.
\end{equation}
The generalised frame field and the projected structure coefficients $F_{\pb I\p A}{}^{\p B}$ are invariant under this transformation. They are rather governed by $\delta^{(0)}_\lambda E_A{}^I = \lambda_A{}^B E_B{}^I$ which implies, due to the frame algebra~\eqref{eqn:framealg},
\begin{equation}
  \delta^{(0)}_\lambda F_{\pb I\p A}{}^{\p B} = \partial_{\pb I} \lambda_{\p A}{}^{\p B} + \lambda_{\p A}{}^{\p C} F^{(0)}_{\pb I \p C}{}^{\p B} + F^{(0)}_{\pb I\p A}{}^{\p C} \lambda^{\p B}{}_{\p C}\,.
\end{equation}
It is important to keep in mind that this transformation does not affect $\Theta_{IA}{}^B$. Hence $\tilde{\delta}_\lambda$ should be understood as an auxiliary transformation whose main purpose is to write the Taylor expansion of $\Delta_\Lambda$ around the identity transformation in the compact form
\begin{equation}
  \Delta_\Lambda = \sum_{n=1}^\infty \frac1{n!}(\left. \tilde{\delta}_\lambda)^n \Delta_\Lambda \right|_{\Theta = 0} = \sum_{n=1}^\infty \frac1{n!}(\delta_\lambda)^n\,.
\end{equation}
By taking into account the definition of a finite transformation as the exponential map of its infinitesimal version, we are able to read off $\delta_\lambda$ directly from the leading contribution of this expansion. Additionally, one has to verify that all subleading contributions match as well. Otherwise, the proposal for $\Delta_\Lambda$ would be inconsistent and should be discarded. Fortunately, both \eqref{eqn:trafoG} and \eqref{eqn:trafoB} satisfy the relation
\begin{equation}
  \tilde \delta_\lambda \Delta^{(0)}_\Lambda - \delta^{(0)}_\lambda \Delta^{(0)}_\Lambda = \delta^{(0)}_\lambda
\end{equation}
that implies $(\tilde{\delta}_\lambda)^n \Delta^{(0)}_\Lambda |_{\Theta=0} = (\delta^{(0)}_\lambda)^n$ and therefore guarantees the correctness of the proposed transformations.

To make contact with the known expressions for $\delta_\lambda$ in the literature, we first calculate
\begin{equation}\label{eqn:deltaRBWZW}
  \tilde\delta_\lambda \mathcal{B}^{\mathrm{WZW}}_{IJ} = \partial_{[I|} \lambda_A{}^B \Theta_{|J]B}{}^A + \partial_{[I} \xi_{J]}
\end{equation}
which is only defined up to a shift by a closed two-form. By the Poincar\'e lemma, this two-form is in local patches exact where it can be parameterised by $\xi_I = ( 0 \,\, \xi_i )$. Equation~\eqref{eqn:deltaRBWZW} implies
\begin{equation}
  \delta^{(0)}_\lambda \mathcal{B}_{IJ} = \partial_{[\pb I|} \lambda_{\p A}{}^{\p B} F^{(0)}_{|\pb J]\p B}{}^{\p A} + (\partial\xi)_{[\pb I\pb J]}
\end{equation}
and ultimately $\delta^{(1)}_\lambda K_{IJ}$. Note that in the last term projections are applied after taking the derivative. This is important because both operations do not commute. Contact with \cite{Marques:2015vua} is made through the infinitesimal transformation of the generalised frame field
\begin{equation}
  \delta^{(1)}_\lambda E^A{}_I E^{(0)}_{AJ} = \partial_{[\p I|} \lambda_{\p A}{}^{\p B} F^{(0)}_{|\pb J]\p B}{}^{\p A} + (\partial \xi)_{[\p I \pb J]}
\end{equation}
where we extended (3.24) by a compensation $B$-field transformation. The Born structure gives rise to $K_I{}^J P_J{}^K = \overline{P}_I{}^J K_J{}^K$, $K_I{}^J \partial_J = \partial_I$ and eventually allows us to establish
\begin{equation}
  \begin{aligned}
    \delta^{(1)}_\lambda K_{IJ} &= 2 \delta^{(1)}_\lambda E^A{}_{[I|} E^{(0)}_A{}^K K^{(0)}_{K|J]} \\
    &= \delta^{(0)}_\lambda \mathcal{B}_{IJ} - K^{(0)}_I{}^K \delta^{(0)}_\lambda \mathcal{B}_{KL} K^{(0)}_J{}^L\,.
  \end{aligned}
\end{equation}
Hence, the finite transformation~\eqref{eqn:doubleLK} is indeed compatible with the known infinitesimal version. The same applies to~\eqref{eqn:doubleLgenmetric}, which we rewrite as
\begin{equation}
  \begin{aligned}
    \Delta^{(1)}_\Lambda \mathcal{H}_{IJ} = 2 K^{(0)}_{(I|}{}^K \Bigl(& \Delta^{(0)}_\Lambda \mathcal{B}_{K|J)} + \Delta^{(0)}_\Lambda \mathcal{G}_{K|J)} \Bigr)
  \end{aligned}
\end{equation}
to find
\begin{equation}
  \delta^{(1)}_\Lambda \mathcal{H}_{IJ} = 2 \partial_{(\p I|} \lambda_{\p A}{}^{\p B} F^{(0)}_{|\pb J)\p B}{}^{\p A} + 2 (\partial \xi)_{(\p I\pb J)}\,.
\end{equation}
We obtain a match with (3.27) of \cite{Marques:2015vua} and conclude our discussion of finite gGS transformations.

The $\alpha'$-corrected PL T-duality transformation rules in the MN scheme arise after~\eqref{eqn:trgenmetric0} is adapted to take into account the non-trivial action of double Lorentz transformations on the generalised metric beyond the leading order in $\alpha'$, namely
\begin{equation}\label{eqn:trgenmetric1}
  \widetilde{\mathcal{H}}^{IJ} = O_K{}^I O_L{}^J \left( \mathcal{H}^{KL} + \Delta^{(1)}_{\widetilde{\Lambda}\Lambda^{-1}} \mathcal{H}^{KL} \right)\,.
\end{equation}
Like $O_I{}^J$ in \eqref{eqn:defO}, the transformation parameter of the gGS transformation, $(\widetilde{\Lambda} \Lambda^{-1})_A{}^B = \widetilde{\Lambda}_A{}^C \Lambda^B{}_C$, is directly extracted from the corresponding generalised frame fields. The generalised dilaton is invariant under this transformation and thus \eqref{eqn:trgend} still applies. This however does not imply that the dilaton $\phi$ is resistant to $\alpha'$-corrections. It depends on both, the generalised dilaton and the determinant of the target space metric, and the latter receives corrections. For completeness, let us note that the four derivative effective action in the MN scheme is given in (3.38) of \cite{Marques:2015vua}. The field redefinitions which are required to go to the gBR and the MT scheme can also be found in this paper (in equations (3.67) and (B.7), respectively). In the presented DFT formulation, it is manifest that~\eqref{eqn:trgenmetric1} will not change the action nor the corresponding field equations, since both can be exclusively written in terms of the structure coefficients $F_{AB}{}^C$ \cite{Baron:2017dvb}. Hence, it is guaranteed that two-loop conformal invariance of a PL symmetric $\sigma$-model is preserved. An important but more subtle question is if this result can be extended to RG flows between CFTs, like it is possible at one loop. Here a significant challenge is that the relation between $\beta$-functions and field equations of the effective action becomes more and more complicated with increasing loop order. However, recently presented $\alpha'$-corrected RG flows for integrable and PL symmetric $\eta$- and $\lambda$-deformations \cite{Hoare:2019ark,*Hoare:2019mcc} suggest that it is possible to overcome this problem in the future.

Another aspect that deserved further investigation is the, at least for us initially surprising, connection to Born geometry. In contrast to Riemannian geometry where the Levi-Civita connection is unique, DFT does not possess a completely determined, torsion-free covariant derivative which is compatible with both, $\eta_{IJ}$ and the generalised metric. Consequentially, the generalised Riemann tensor contains undetermined contributions \cite{Hohm:2011si}. They drop out in all physically relevant quantities at the two derivative level, like the generalised Ricci scalar and tensor. However, it is not possible to construct the Riemann tensor squared term that captures $\alpha'$-corrections of the effective action directly from the generalised Riemann tensor. Born geometry already was argued to help to obtain a unique connection by additionally requiring compatibility with $K$ \cite{Freidel:2018tkj}. Considering our results, one might hope that it also gives valuable insights into the generalised geometry of $\alpha'$-corrections.

\paragraph*{Note added:} During the completion of this letter, we learned of the closely related independent work \cite{Borsato:2020bqo2}.

\begin{acknowledgments}
\paragraph*{Acknowledgements:}
We thank Daniel Butter and William Linch for inspiring discussions and Christopher Pope for comments on the draft.
\end{acknowledgments}

\bibliographystyle{apsrev4-1}
\bibliography{literature}

\end{document}